\documentclass[a4j,11pt]{article}

\usepackage{amsmath,amssymb,color,graphics,amscd,amsfonts,epsf}
\allowdisplaybreaks[4]

\date{}
\begin{document}

\begin{flushright} 
 
WIS/10/09-JULY-DPP\\
APCTP Pre2009 - 010

\end{flushright} 

\vspace{0.1cm}

\begin{center}
  {\LARGE
  
   Large-$N$ reduced models of supersymmetric 

quiver, Chern-Simons gauge theories and ABJM
    
  }
\end{center}
\vspace{0.1cm}
\vspace{0.1cm}
\begin{center}

         Masanori H{\sc anada}$^{a}$\footnote
         {
E-mail address : masanori.hanada@weizmann.ac.il},  
         Lorenzo M{\sc annelli}$^{a}$\footnote
         {
E-mail address : lorenzo.mannelli@weizmann.ac.il} 
and 
         Yoshinori M{\sc atsuo}$^{b}$\footnote
         {
E-mail address : ymatsuo@apctp.org}         

\vspace{0.3cm}

${}^{a}$
{\it Department of Particle Physics, Weizmann Institute of Science,\\
     Rehovot 76100, Israel }\\

${}^{b}$
{\it Asia Pacific Center for Theoretical Physics,\\
Pohang, Gyeongbuk 790-784, Korea}\\
\end{center}


\vspace{1.5cm}

\begin{center}
  {\bf Abstract}
\end{center}

Using the Eguchi-Kawai equivalence, 
we provide regularizations of supersymmetric quiver and Chern-Simons gauge theories  
which leave the supersymmetries unbroken. 
This allow us to study many interesting  
theories on a computer. As examples we construct large-$N$ reduced models of  
supersymmetric QCD with flavor and the ABJM model of multiple M2 branes. 

\newpage
\newpage

\section{Introduction}
\hspace{0.51cm}
Super Yang-Mills theories (SYM) have attracted much interests 
as leading candidates for physics beyond the standard model. 
To understand their nonperturbative aspects, 
one might expect lattice simulation to be useful 
as it happens in the usual non-supersymmetric gauge theories.   
However, it is generally difficult to realize SYM on lattices and consequently 
detailed numerical simulations have been performed only in 
lower (less than four) dimensions (see 
\cite{4dN=1LatticeSimulation} for recent simulations of 4d SYM, and 
\cite{CKU09} for a review of lattice supersymmetry).  

SYM's are also important because it is expected to provide a nonperturbative 
formulation of superstring/M theory at large-$N$ 
\cite{BFSS96,IKKT96,DVV97,Maldacena97,IMSY98}.  
In this context the relevant theories are SYM's in lower dimensional spacetime. 
In particular, $(0+1)$-dimensional theory can be analyzed on computer 
by using the non-lattice technique \cite{HNT07} 
(in this case lattice simulations are also possible \cite{CW07}) 
and a part of these conjectures has been confirmed by Monte-Carlo simulations 
in the strong coupling regime \cite{AHNT07,CW08,HNT08}.  
Chern-Simons gauge theories are also relevant in the context 
of superstring/M theory. 
In fact recently supersymmetric Chern-Simons gauge theories in three dimensions 
have been proposed as a description of the theory of 
multiple M2-branes \cite{Schwarz04,ABJM08}. 
Since these theories describe membranes in their strong coupling regimes, 
numerical simulations appear to be a very useful tool. 
Unfortunately it turns out to be a very difficult task to study either 
Chern-Simons or supersymmetry on a lattice. 

At large-$N$, it is possible to circumvent the lattice-SUSY problem 
by using the Eguchi-Kawai equivalence \cite{EK82}. 
This construction guarantees that the large-$N$ gauge theories are equivalent to 
the lower dimensional matrix models if a certain condition is satisfied. 
Recently this equivalence has been used to formulate 4d ${\cal N}=4$ SYM 
compactified on $S^3$ \cite{IIST08}.%
\footnote{
For other attempts that use the Eguchi-Kawai equivalence, see 
\cite{KUY07,Bringoltz09,BBCS09}.} 
In this construction, the BMN matrix model \cite{BMN02} 
around a certain multi-fuzzy sphere solution 
is argued to be equivalent, using the Eguchi-Kawai reduction, 
to 4d ${\cal N}=4$ SYM. 
(For evidence supporting the validity of this formulation, see 
\cite{IKNT08}.) 
Given that the solution is BPS, it provides 
a regularization that preserve part of supersymmetry. 
Furthermore, given that one-dimensional system like the BMN model 
can be analyzed on computer, the previous formulation allows us to study 
4d SYM using a Monte-Carlo simulation. 

A natural question to ask is to which kind of theories it is possible to apply
the Eguchi-Kawai equivalence. 
The generalization to 4d ${\cal N}=1$ pure SYM is 
quite straightforward, as discussed in \cite{HMM09},
using the matrix model analyzed in \cite{KP06}. 
However, introducing fundamental matter, as will be explained in 
\S~\ref{section:Quiver}, turns out to be a difficult task\footnote{
As discussed in \cite{ASV03}, large-$N$ gauge theories with 
a quark in the two-index antisymmetric representation can be regarded 
as the counterpart of the usual QCD, in the sense that this representation 
reduces to the fundamental representation when $N=3$. 
For these models, Eguchi-Kawai reduction can easily applied. See e.g. \cite{Bringoltz09}. 
}.   
In this paper, we show that quiver and 
Chern-Simons gauge theories can be regularized 
using the techniques of \cite{IIST08}. 
More specifically, in \S~\ref{section:Quiver} 
we construct the supersymmetric $SU(N)\times SU(M)$ gauge theory 
with bifundamental matter. 
Then, by sending the coupling constant of the latter gauge group to zero, 
we obtain a global flavor symmetry from this gauge symmetry, and 
as a consequence, this quiver gauge theory becomes supersymmetric QCD. 
In this construction both $N_c=N$ and $N_f=MK$ ($K$ is the number of 
bifundamental matters) must be infinite, but the ratio $N_f/N_c$ 
can be arbitrary. 
In \S~\ref{section:CS} we show that Chern-Simons theory, 
which is difficult to 
study on lattice, can also be formulated in terms of Eguchi-Kawai equivalence. 
For that purpose, we use a construction of the Chern-Simons theory 
using a generalization of Taylor's T-duality prescription \cite{Taylor96} 
which is discussed in \cite{IIOST07}. 
Combining this results with the technique of \cite{IIST08}  
the Eguchi-Kawai formulation can be obtained straightforwardly.
\footnote{
While this work was in progress we have been informed that the same idea 
had been studied by Ishiki et al \cite{IOST09}. 
We thank G.~Ishiki for the discussion. 
} 
Furthermore combining the analysis for the quiver and Chern-Simons theories, 
we are able to construct the ABJM theory \cite{ABJM08} from a matrix model. 

This paper is organized as follows. 
In \S~\ref{section:EguchiKawaiModel} 
we review the basic ideas of the Eguchi-Kawai equivalence. 
First, in \S~\ref{sec:QEK}, we explain the quenched Eguchi-Kawai model 
\cite{BHN82,Parisi82}.  
Based on it, in \S~\ref{sec:3d example}, 
we review the reduced model of SYM on $S^3$ \cite{IIST08}.  
In \S~\ref{section:Quiver} we generalize 
this technique to construct supersymmetric quiver gauge theories. 
In \S~\ref{section:CS} 
we formulate Chern-Simons theory in three dimensions 
along the line of \cite{IIOST07}. 
Combining these results with the ones in 
\S~\ref{section:Quiver} we construct the Chern-Simons-matter theories 
which  recently attracted much interest as the theory 
describing multiple M2-branes. 
\section{The basics of the Eguchi-Kawai reduction}\label{section:EguchiKawaiModel}
\hspace{0.51cm}
In this section we review the Eguchi-Kawai equivalence \cite{EK82}. 
In \S\ref{sec:QEK} we introduce the ``quenched'' version of the Eguchi-Kawai model 
\cite{BHN82,Parisi82}, which is relevant for our purpose.  
In \S~\ref{sec:3d example} we use this technique to formulate 
large-$N$ Yang-Mills on the three-sphere. 
\subsection{Quenched Eguchi-Kawai model}\label{sec:QEK}
\hspace{0.51cm}
In the following we review the diagrammatic approach to 
the quenched Eguchi-Kawai model(QEK) \cite{Parisi82}. 
The basic idea is that 
{\it in the planar limit, Yang-Mills theory is equivalent to 
a matrix model around a suitable background}. 
We will also consider QEK for compact space \cite{IIST08, KS07}. 
In order to see clearly the difference between the compact 
and noncompact cases, 
we consider (analogously to \cite{IIST08}) the simplest example first, 
namely the correspondence between a zero-dimensional matrix model 
and a matrix quantum mechanics.  

As a simple example, we consider a matrix quantum mechanics with a mass term, 
\begin{eqnarray}
S_{1d}=N\int dt Tr\left(
\frac{1}{2}(D_t X_i)^2
-
\frac{1}{4}[X_i,X_j]^2
+
\frac{m^2}{2} X_i^2
\right),  
\end{eqnarray}
where $X_i\ (i=1,2,\cdots,d)$ are $N\times N$ traceless Hermitian matrices. The covariant derivative $D_t$ is 
given by 
\begin{eqnarray}
D_t X_i
=
\partial_t X_i - i[A,X_i]. 
\end{eqnarray}
At large-$N$, this model can be reproduced starting from the zero-dimensional model
\begin{eqnarray}
S_{0d}
=
\frac{2\pi}{\Lambda}
\cdot N
Tr\left(
-
\frac{1}{2}[Y,X_i]^2
-
\frac{1}{4}[X_i,X_j]^2
+
\frac{m^2}{2} X_i^2
\right), 
\label{0d action}
\end{eqnarray}
where $Y$ and $X_i$ are $N\times N$ traceless Hermitian matrices. 
We embed the (regularized) translation generator into 
the matrix $Y$, 
\begin{eqnarray}
Y^{b.g.}=diag(p_1,\cdots,p_N), 
\qquad
p_k=\frac{\Lambda}{N}\left(
k-\frac{N}{2}
\right). 
\end{eqnarray}
By expanding $Y$ around this background,  
\begin{eqnarray}
Y
=
Y^{b.g.}+A,   
\end{eqnarray}
the Feynman rules of the one-dimensional theory, as we will see in the following, are reproduced at large-$N$.  

\begin{figure}[tbp]
\begin{center}
\scalebox{0.5}{
\rotatebox{0}{
\includegraphics*{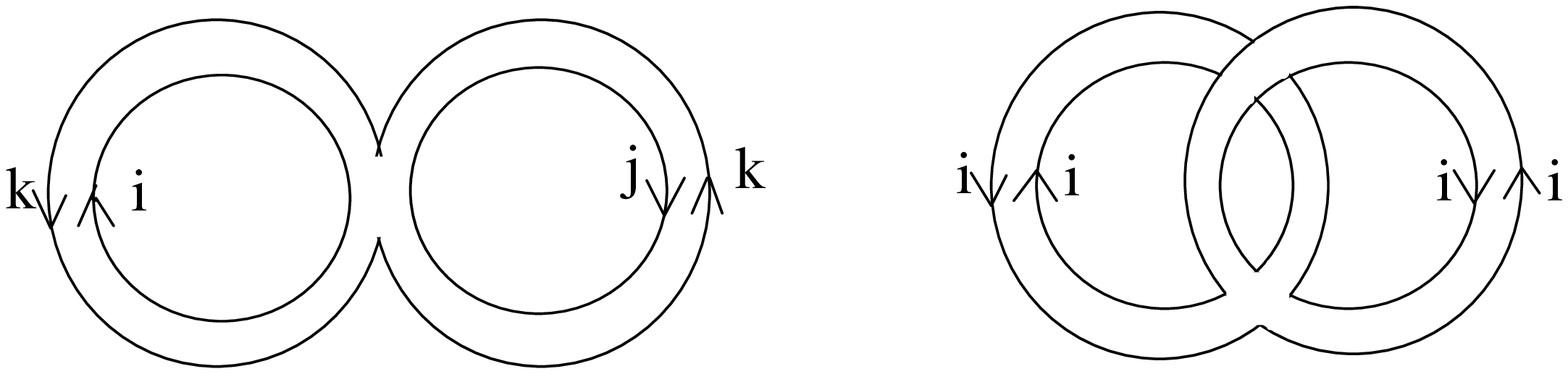}
}}
\caption{Two-loop planar and nonplanar diagrams with quartic interaction vertex. 
}\label{fig:TwoLoop}
\end{center}
\end{figure}

The action can be rewritten as 
\begin{eqnarray}
S_{0d}
=
\frac{2\pi}{\Lambda}
\cdot N
\Biggl\{
\frac{1}{2}
\sum_{i,j}\left|
(p_i-p_j)\left(X_k\right)_{ij}
-
i[A,X_k]_{ij}
\right|^2
+
Tr\left(
-
\frac{1}{4}[X_i,X_j]^2
+
\frac{m^2}{2} X_i^2
\right)
\Biggl\}. 
\end{eqnarray}
We add to it the gauge-fixing and Faddeev-Popov terms 
\begin{eqnarray}
\frac{2\pi}{\Lambda}
\cdot N
Tr\left(
\frac{1}{2}[Y^{b.g.}, A]^2
-
[Y^{b.g.},b][Y,c]
\right).  
\end{eqnarray}
Then, the planar diagrams are the same as the ones in the 1d theory up to a normalization factor. 
For example, a scalar two-loop planar diagram with quartic interaction 
(see Fig.\ref{fig:TwoLoop}) is 
\begin{eqnarray}
\lefteqn{
\frac{d(d-1)}{2}
\left(
\frac{1}{2}\cdot\frac{2\pi N}{\Lambda}
\right)
\sum_{i,j,k=1}^N
\frac{(\Lambda/2\pi N)}{(p_i-p_k)^2+m^2}
\frac{(\Lambda/2\pi N)}{(p_j-p_k)^2+m^2}
}\nonumber\\
&\simeq&
\frac{d(d-1)}{4}\cdot\frac{2\pi}{\Lambda}\cdot
N^2
\int^{\Lambda/2}_{-\Lambda/2} \frac{dp}{2\pi}
\int^{\Lambda/2}_{-\Lambda/2} \frac{dq}{2\pi}
\frac{1}{(p^2+m^2)(q^2+m^2)}.   
\end{eqnarray}
The essence of this expression is that {\it the adjoint action of 
the background matrix can be identified with the derivative} 
and
{\it the matrix elements of the fluctuations can be identified with the Fourier modes 
in momentum space}. 
The corresponding diagram in the 1d theory is 
\begin{eqnarray}
\frac{d(d-1)}{4}\cdot Vol
\cdot N^2
\int^{\Lambda/2}_{-\Lambda/2} \frac{dp}{2\pi}
\int^{\Lambda/2}_{-\Lambda/2} \frac{dq}{2\pi}
\frac{1}{(p^2+m^2)(q^2+m^2)}, 
\end{eqnarray}
where $Vol$ is volume of the spacetime. 
Hence by interpreting $\Lambda$ and $\Lambda/N$ to be UV and IR cutoffs, 
those diagrams agree up to the factor $\left(\frac{\Lambda}{2\pi}\right)\cdot Vol$. 
The other planar diagrams also correspond up to the same factor. 

The nonplanar diagrams do not have such a correspondence, but in an appropriate limit 
they are negligible. In the 1d theory, by taking a planar limit 
they are suppressed by a factor $N^{-2}$. In the reduced model, they are suppressed 
if IR cutoff $\Lambda/N$ goes to zero. To see this, let us calculate for example the 
two-loop nonplanar diagram in Fig.\ref{fig:TwoLoop}. 
It reads 
\begin{eqnarray}
-\frac{d(d-1)}{4m^4}\frac{\Lambda}{2\pi}, 
\end{eqnarray}  
which is suppressed by a factor $(\Lambda/N)^2$ compared with planar diagrams. 

Therefore, by taking the limit 
\begin{eqnarray}
N\to\infty, 
\qquad
\Lambda\to\infty, 
\qquad
\frac{\Lambda}{N}\to 0
\end{eqnarray}
the 1d model on ${\mathbb R}$ is reproduced from the 0d model. 

If one wants to obtain the theory on a circle, it is necessary to 
fix the IR cutoff while suppressing nonplanar diagrams. This can be achieved 
by taking the background to be  
\begin{eqnarray}
Y^{b.g.}=diag(p_1,\cdots,p_{n_1})\otimes \textbf{1}_{n_2}, 
\qquad
p_k=\frac{\Lambda}{n_1}\left(
k-\frac{n_1}{2}
\right), 
\qquad
N=n_1n_2,  
\end{eqnarray}
and taking $n_1,n_2$ and $\Lambda$ to be infinity 
while fixing the IR cutoff $\Lambda/n_1$ \cite{IIST08}.%
\footnote{
Another reason to consider the $n_2\to\infty$ limit is the following,
if we take $n_2=1$ and quench the background, no zero-mode appears. 
This is not a problem when we take the noncompact limit, because 
the IR cutoff goes to zero. On the other hand, when we consider a  compact space, 
the absence of the zero-mode destroys the Eguchi-Kawai equivalence. 
It turns out that by taking the $n_2\to\infty$ limit, zero-modes are taken into account in an appropriate manner.   
} 

It turns out that in this setup the background is not stable. 
So, to make the expansion meaningful, we have to ``quench'' the eigenvalues of $Y$, i.e. we have to fix the background by hand. 
This is the reason for the name ``quenched'' Eguchi-Kawai model. 
\subsection{Eguchi-Kawai construction of Yang-Mills on $S^3$}\label{sec:3d example}
\hspace{0.51cm}
Next let us construct the Yang-Mills theory on the three-sphere 
by using the Eguchi-Kawai equivalence. 
The essence of QEK is to find a background whose adjoint action can be identified 
with the spacetime derivative. So, the strategy is 
{\it to find a set of three matrices 
whose adjoint action can be identified with the derivative on $S^3$}. 
Such matrices were found in \cite{ISTT06,IIST08}. 
In the following we will show the derivation in a heuristic way.  
\subsubsection{YM on $S^3$}
\hspace{0.51cm}
In this section, we express the action of Yang-Mills theory  
on $\mathbb R \times S^3$ in a form convenient for our purpose \cite{IIST08}.
The radius of the sphere is taken to be $2/\mu$.
The action of $U(N)$ SYM is given by 
\begin{eqnarray}
S
=
-\frac{N}{\lambda_{4d}}\int dt\int_{S^3} d^3x \sqrt{-g(x)}
Tr\frac{1}{4}F_{\mu\nu}^2,    
\end{eqnarray}
where $\lambda_{4d}$ is the 't Hooft coupling constant, $g_{\mu\nu}(x)$ 
is the metric and $g(x)$ is its determinant. 
The field strength is 
\begin{eqnarray}
F_{\mu\nu}
=
\partial_\mu A_\mu
-
\partial_\nu A_\mu
-
i[A_\mu,A_\nu].    
\end{eqnarray} 
The Greek indices $\mu$, $\nu$ refer to 
the Einstein frame and the Latin indices to the local Lorentz frame. 

The sphere part of this geometry has the group structure of $SU(2)$. 
Given this group structure, 
there exists a right-invariant 1-form $dg g^{-1}$ 
and its dual Killing vectors ${\cal L}_i$, 
satisfying the commutation relation 
\begin{eqnarray}
[{\cal L}_i,{\cal L}_j]=i\epsilon_{ijk}{\cal L}_k. 
\end{eqnarray}
Using the coordinates $(\theta,\psi,\varphi)$ defined by 
$g=e^{-i\varphi\sigma_3/2}e^{-i\theta\sigma_2/2}e^{-i\psi\sigma_3/2}$, 
the vielbein $E^i$ can be expressed as 
\begin{eqnarray}
 E^1 
  &=& 
  \frac{1}{\mu}
  \left(
   -\sin\varphi d\theta + \sin\theta \cos\varphi d\psi 
  \right) , \\
 E^2 
  &=& 
  \frac{1}{\mu}
  \left(
   \cos\varphi d\theta + \sin\theta \sin\varphi d\psi    
  \right) , \\
 E^3 
  &=& 
  \frac{1}{\mu}
  \left(
   d\varphi + \cos\theta d\psi 
  \right) . 
\end{eqnarray}
In these coordinates the metric is given by 
\begin{equation}
 ds^2 = \frac{1}{\mu^2}
  \left[
   d\theta^2 + \sin^2\theta\,d\varphi^2 
   + \left(d\psi + \cos\theta\,d\varphi\right)^2
  \right] . 
\end{equation}
The spin connection $\omega_{abc}$ can be read off 
from the Maurer-Cartan equation, 
\begin{eqnarray}
 dE^i - \omega^i_{\ jk} E^j \wedge E^k &=& 
 0 , \\
 \omega_{ijk} &=& \frac{\mu}{2}\epsilon_{ijk} . 
\end{eqnarray}
and the Killing vectors are given by 
\begin{eqnarray}
{\cal L}_i=-\frac{i}{\mu}E_i^M\partial_M, 
\end{eqnarray}
where
\begin{eqnarray}
{\cal L}_1
&=&
-i\left(
-\sin\varphi\partial_\theta
-
\cot\theta\cos\varphi\partial_\varphi
+
\frac{\cos\varphi}{\sin\theta}\partial_\psi
\right), 
\nonumber\\
{\cal L}_2
&=&
-i\left(
\cos\varphi\partial_\theta
-
\cot\theta\sin\varphi\partial_\varphi
+
\frac{\sin\varphi}{\sin\theta}\partial_\psi
\right), 
\nonumber\\
{\cal L}_3
&=&
-i\partial_\varphi.
\label{SU(2) generator on S^3} 
\end{eqnarray} 
The Killing vectors represent a  complete basis for the tangent space on $S^3$. 
Furthermore given that  the vielbeins are defined everywhere on $S^3$, 
the indices $i$ can be used as a label for
the vector fields and 1-forms.%
\footnote{
This property is necessary in order to identify this index with 
the one in the matrix model \cite{HKK05}. 
} 

By using the Killing vectors ${\cal L}_i$, 
the action can be rewritten as \cite{IIST08}
\begin{eqnarray}
S 
&=&
\left(
\frac{2}{\mu}
\right)^3
\frac{N}{\lambda_{4d}}\int dt\int d\Omega_3 Tr
\Biggl(
\frac{1}{2}
\left(
D_t A_i
-
\mu{\cal L}_iA_t
\right)^2
\nonumber\\ 
& &
\qquad\qquad
+
\frac{\mu^2}{4}({\cal L}_iA_j-{\cal L}_jA_i)^2
-
\frac{\mu}{2}({\cal L}_iA_j-{\cal L}_jA_i)[A_i,A_j]
+
\frac{1}{4}[A_i,A_j]^2
\nonumber\\
& &
\qquad\qquad
-
\frac{\mu^2}{2}A_i^2
+
i\mu\epsilon^{ijk}A_iA_jA_k
-
i\mu^2\epsilon^{ijk}A_i({\cal L}_jA_k)
\Biggl) , 
\label{YM action in Maurer-Cartan basis}
\end{eqnarray} 
where $A_i$ is defined in such a way that the 1-form of the gauge field 
on $S^3$ take the form $A = A_i E^i$, and 
$d\Omega_3$ is the volume form of the unit three-sphere.  
\subsubsection{Eguchi-Kawai reduction}
\hspace{0.51cm}
To construct matrices which represent derivatives on $S^3$ 
in a coordinate-independent way, 
it is useful to use the $SU(2)$ group structure of $S^3$.  
The Killing vectors \eqref{SU(2) generator on S^3} 
act on functions on $S^3\simeq SU(2)$, whose irreducible decomposition is%
\footnote{
In general, for a compact Lie group $G$, a space of functions on $G$ is decomposed as 
\begin{eqnarray}
{\cal C}^\infty (G)
=
\bigoplus_{r}
\left(
\underset{d_r-{\rm times}}{
\underbrace{
V_r\oplus\cdots\oplus V_r}}
\right), 
\end{eqnarray}
where $r$ runs over all the irreducible representations, $V_r$ is a representation space 
and $d_r$ is its dimension. 
} 
\begin{eqnarray}
{\cal C}^\infty (SU(2))
=
\bigoplus_{J=0,1/2,1,\cdots}
\left(
\underset{(2J+1)\text{-times}}{
\underbrace{
V_J\oplus\cdots\oplus V_J}}
\right), 
\end{eqnarray}
where $V_J$ is the space that the spin $J$ representation acts on. 

In order to realize this representation as the adjoint action of the background matrices, 
we first embed the $SU(2)$ generators into $N\times N$ matrices. 
We then introduce the matrices $L_i$ which satisfy 
the commutation relation of the $SU(2)$ generators, 
\begin{equation}
 [L_i , L_j ] = i \epsilon_{ijk} L_k . \label{SU(2) commutation}
\end{equation}
Since these matrices cannot be diagonalized simultaneously, 
we embed them in the following block diagonal form; 
\begin{eqnarray}
L_i
=
\left(
\begin{array}{ccccc}
  \ddots & & & & \\
  & L_i^{[j_{s-1/2}]}& & & \\
  & & L_i^{[j_s]} & & \\
  & & & L_i^{[j_{s+1/2}]} & \\
  & & & & \ddots
  \end{array}
\right),  \label{IIST background}
\end{eqnarray}
where $L_i^{[j_s]}$ is a $(2j_s+1)\times (2j_s+1)$ matrix 
which acts on the spin $j_s$ representation. 
The size of the matrix $N$ is 
\begin{eqnarray}
 N=\sum_s (2j_s+1) . 
\end{eqnarray}
We introduce a regularization by restricting 
the representation space to a limited number of $j_s$. 
Furthermore we take the integer $s$ satisfying
\begin{eqnarray}
-\frac{T}{2}\le s\le \frac{T}{2}, 
\end{eqnarray} 
where $T$ is an even integer. 
We introduce another integer $P \gg T$ 
and take $j_s$ to be
\begin{eqnarray}
j_s
=
\frac{P+s}{2}.   
\end{eqnarray}
The large $N$ limit is taken in the following way
\begin{eqnarray} 
P
\to\infty, 
\qquad
T\to\infty, 
\qquad
N\to\infty. 
\end{eqnarray} 

To see how this prescription works, let us consider the $(j,j')$-block, to which 
$L^{[j]}$ acts from left and $L^{[j']}$ acts from the right. 
A Basis for this block is symbolically written as  
\begin{eqnarray}
|j,m\rangle\langle j',m'|.  
\end{eqnarray} 
It can be decomposed into spin $|j-j'|,\cdots,j+j'$ representations. 
Let's count the number of representations of each spin.  
\\

\begin{tabular}{r@{Spin }c@{ : }l} 
 & 0 & $T+1$, because it appears only when $j=j'\ge 0$. \\[6pt]
 & 1/2 & $2T$, because it appears when $j=j'\pm 1/2$. \\[6pt]
 & 1 & $(T+1)+2(T-1)=3T-1$, 
 because it appears when $j=j'\ge 1$ and $j=j'\pm 1$. \\[6pt]
 & $J\in{\mathbb Z}$ & $(T+1)+\sum_{l=1}^J 2(T+1-2l)=(2J+1)T+1-2J^2$. 
\end{tabular}\\

As long as $T\gg J$, we can approximate this expression as 
\begin{eqnarray}
({\rm number\ of\ spin\ }J)
\simeq
(2J+1)T. 
\end{eqnarray} 
Therefore the representation space, or equivalently the variables appearing 
in the matrix model, can be regarded as a set of $T$ copies of the space of functions on $S^3$. 
As $J$ increase the number of copies decreases. In this sense $T$ plays a role of a momentum cutoff.

In this way matrix elements can be identified with the functions on $S^3$, 
or in other words the propagators in the Feynman diagram agrees.  
However it is not apparent if this identification is consistent with the multiplication 
of the fields. (This is necessary in order for the interaction vertices to agree.) 
In \cite{IIST08} it has been shown that 
\begin{eqnarray}
\frac{T}{P}\to 0
\end{eqnarray} 
is a sufficient condition for the compatibility with the multiplication.%
\footnote{
Given that this condition force us to use very large matrix 
in a computer simulation, 
it would be nice if it could be relaxed. 
However this seems to be impossible because the
eigenvalue distribution is not uniform without imposing $T/P\to 0$, 
while the eigenvalues should be distributed uniformly in order for 
the quenched Eguchi-Kawai equivalence to work. 
One possible solution is to make the density uniform by putting  
fuzzy spheres with the same radii on top of each other and tune the number 
of copies to be proportional to the spin. 
We thank G.~Ishiki for stimulating discussion on this point. 
} 

By using these matrices we can relate 
a matrix model to a gauge theory on $S^3$, given by the action 
(\ref{YM action in Maurer-Cartan basis}). 
In order to do that, we consider the bosonic matrix quantum mechanics 
\begin{eqnarray}
S=
C\cdot \frac{N}{\lambda_{4d}}\int dt Tr\left(
\frac{1}{2}(D_t X_i)^2
+
\frac{1}{4}[X_i,X_j]^2 
+
i\mu\epsilon_{ijk}X^iX^jX^k 
-
\frac{\mu^2}{2}X_i^2
\right),   
\end{eqnarray}
where the constant $C$ will be specified shortly.  
We then expand the action around a classical solution 
\begin{eqnarray}
A_t=0,\qquad X_i=-\mu L_i,  
\end{eqnarray}
identify ${\cal L}_i$, $A_t$ and $A_i$ with 
$[L_i,\ \cdot\ ]$, $A_t$ and $X_i + \mu L_i$, 
\begin{eqnarray}
{\cal L}_i
\to
[L_i,\ \cdot\ ],
\qquad 
A_t^{(4d)}
\to
A_t^{(1d)}, 
\qquad
A_i
\to 
X_i + \mu L_i, 
\end{eqnarray}
and replace the trace and the spatial integral by a trace, 
\begin{eqnarray}
\left(\frac{2}{\mu}\right)^3\int d\Omega_3 Tr
\to
Tr. 
\end{eqnarray}
The UV and IR momentum cutoffs are given by $\mu T$ and $\mu$, respectively, 
and we will take the limit in such a way that 
\begin{eqnarray}
\mu\to 0, 
\qquad
\mu T\to \infty. 
\end{eqnarray}
In order to match the diagrams completely, the coupling constant should be taken as \cite{IIST08}
\begin{eqnarray}
\lambda_{4d}
=
\lambda_{1d}\cdot\frac{16\pi^2}{\mu P}.   
\label{correspondence_coupling}
\end{eqnarray}
In other words, we have to multiply the dimensionally reduced action 
by an overall factor
\begin{eqnarray}
C\equiv\frac{16\pi^2}{\mu P} . 
\end{eqnarray}
This factor is analogous to 
the factor $2\pi/\Lambda$ in (\ref{0d action}). 
Furthermore the four-dimensional 't Hooft coupling $\lambda_{4d}$ should be scaled with 
the UV momentum cutoff $\mu T$. 

Finally we would like to add a few remarks. 
First, the background is a classical solution and hence as long as 
it is stable we do not need to quench it. 
Second, when we take the large-$N$ limit fixing the IR momentum cutoff $\mu$, 
in order to suppress the nonplanar diagrams it is necessary to change the background to 
\begin{eqnarray}
 - \mu L_i\otimes \textbf{1}_k
\label{multiple IIST background}
\end{eqnarray}
and take $k\to\infty$ limit. 
Thirdly, this construction resembles the ``twisted'' Eguchi-Kawai model (TEK) 
\cite{GAO82}. 
In both cases the model is deformed by background flux terms so that   
noncommutative manifolds (fuzzy sphere for the former and fuzzy torus 
for the latter) become a classical solution, and the higher-dimensional theories 
are obtained as a fluctuation around these solutions. 
\subsection{Supersymmetry and stability of the background} 
\hspace{0.51cm}
So far we have discussed only bosonic theories. 
Strictly speaking the Eguchi-Kawai equivalence does not work 
in bosonic models because the background is unstable. 
This problem arises quite generally, not only in the QEK \cite{BS08}, 
but also in the original reduction \cite{BHN82} and in the TEK \cite{AHHI07,TV06,BNSV06}. 
As usual, supersymmetry can remove such an instability \cite{AHH08}.%
\footnote{
An attempt to avoid the instability 
in the bosonic framework can be found in \cite{UY08}. 
} 
The advantage of the construction explained in \S~\ref{sec:3d example} 
is that supersymmetry can be preserved manifestly \cite{IIST08}. 
That is, the reduced model is a supersymmetric matrix model and 
the background (\ref{IIST background}) is a BPS solution. 
Therefore we can expect the background to be stable at least at low temperature. 

For another approach to the Eguchi-Kawai reduction in supersymmetric Yang-Mills 
with unitary variables, see \cite{Bringoltz09}.  
\section{The Eguchi-Kawai model for quiver gauge theories}\label{section:Quiver}
\hspace{0.51cm}
In \cite{HMM09} 4d ${\cal N}=1$ SYM without flavor is formulated 
by using the Eguchi-Kawai construction of \cite{IIST08}. 
In order to consider QCD, we have to introduce flavors into this model. 
However it is difficult to describe fundamental matter along this line. 
The reason is the following. Because the derivative on the sphere is 
identified to the commutator with matrix $i\mu L_i$, 
the covariant derivative acting on the fundamental 
scalar $\psi$ can be written as 
\begin{eqnarray}
D_i\psi
\sim
i[\mu  L_i,\psi]-iA_i\psi
=
i(\mu L_i-A_i)\psi
-
\psi\cdot i\mu L_i
\equiv
- iX_i\psi  
-
\psi\cdot i\mu L_i .  
\end{eqnarray}
The gauge field $A_i$ acts only from the left, 
and $L_i$ acting from the left can be identified with 
the background of corresponding field in matrix model, $X_i$.
However, the last term in the right hand side cannot be expressed 
as a matrix variable, 
since there are no field acting on $\psi$ from the right. 
(In other words it does not appear from the dimensionally reduced model). 
To circumvent this problem, we consider bifundamental matter. 
Then the covariant derivative becomes 
\begin{eqnarray}
D_i\psi
\sim
i[\mu  L_i,\psi]-iA_i\psi+i\psi B_i
=
i(\mu L_i-A_i)\psi
-
\psi\cdot i(\mu L_i-B_i)
\equiv
- iX_i\psi  
+ 
i\psi Y_i .  
\end{eqnarray}
In this case, both $L_i$ can be identified with 
the background of matrix variables 
and hence the technique of \cite{IIST08} can be applied. 
By taking the additional gauge coupling to be small, 
the gauge field $B_i$ decouples 
and we restore the fundamental matter. 
\subsection{Quiver matrix quantum mechanics and its quenched reduced model}
\hspace{0.51cm}
As the simplest example let us start by considering the bosonic quiver quantum mechanics 
with gauge group $SU(N)\times SU(M)$, where $N$ and $M$ are taken to be infinity 
by fixing the ratio $M/N$. For simplicity we take $M=km$ and $N=kn$, where 
$k,m,n$ are integers, and then, we take the $k\to\infty$ limit fixing $m$ and $n$. 
We consider the following action 
\begin{eqnarray}
S
=
k \int dt Tr \left\{
(D_t\phi)(D_t\phi)^\dagger
+
\mu^2\phi\phi^\dagger
+
g(\phi\phi^\dagger)^2
\right\},  
\end{eqnarray}
where $\phi$ is $N\times M$ matrix and the covariant derivative acts on it as 
\begin{eqnarray}
D_t \phi
=
\partial_t\phi -iA\phi + i\phi B.  
\end{eqnarray}
Here $A$ and $B$ are gauge fields 
associated with $SU(N)$ and $SU(M)$, respectively.  
In this action, the field $\phi$ is rescaled 
such that scaling parameter in $k\to\infty$ appears 
only in the overall factor, furthermore the parameters $\mu$ and $g$ 
do not scale in this limit. 

This model is related to the reduced one
via Eguchi-Kawai equivalence. 
The reduced model is given by 
\begin{eqnarray}
\frac{2\pi}{\Lambda}\cdot k\ Tr
\left\{
-(X\phi - \phi Y)(Y\phi^\dagger - \phi^\dagger X)
+
\mu^2\phi\phi^\dagger
+
g(\phi\phi^\dagger)^2
\right\}. 
\end{eqnarray}
If we expand this action around 
\begin{eqnarray}
X^{b.g.}
=
diag(p_1,\cdots,p_k)\otimes\textbf{1}_n, 
\qquad
Y^{b.g.}
=
diag(p_1,\cdots,p_k)\otimes\textbf{1}_m,
\qquad
p_r
=
\frac{\Lambda}{k}\left(r-\frac{k}{2}\right) , 
\end{eqnarray}
then this model reproduce the results of the original one. 

\begin{figure}[tbp]
\begin{center}
\scalebox{0.5}{
\rotatebox{0}{
\includegraphics*{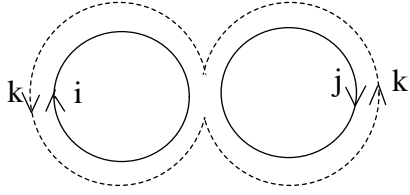}
}}
\caption{A two-loop planar diagram of a bifundamental scalar $\phi$. 
Solid and dotted lines represent $SU(N)$ and $SU(M)$ indices, respectively.  
}\label{fig:TwoLoopQuiver}
\end{center}
\end{figure}
As an example, consider the two-loop diagram shown in Fig.~\ref{fig:TwoLoopQuiver}.  
In the matrix quantum mechanics, it gives 
\begin{eqnarray}
Vol
\cdot k^2\cdot gn^2m
\int^{\Lambda/2}_{-\Lambda/2} \frac{dp}{2\pi}
\int^{\Lambda/2}_{-\Lambda/2} \frac{dq}{2\pi}
\frac{1}{(p^2+\mu^2)(q^2+\mu^2)}, 
\end{eqnarray}
while in the reduced model it is 
\begin{eqnarray}
\frac{2\pi}{\Lambda}
\cdot k^2\cdot gn^2m
\int^{\Lambda/2}_{-\Lambda/2} \frac{dp}{2\pi}
\int^{\Lambda/2}_{-\Lambda/2} \frac{dq}{2\pi}
\frac{1}{(p^2+\mu^2)(q^2+\mu^2)}. 
\end{eqnarray}
Therefore we can see the correspondence as in \S~\ref{sec:QEK}, up to the 
same factor $(\Lambda/2\pi)\cdot Vol$. The generalization to other diagrams is straightforward. 
\subsection{Bosonic quiver gauge theory in four dimensions}
\hspace{0.51cm}
Let us start by considering the bosonic quiver gauge theory with $SU(N)\times SU(M)$ gauge group. 
As in the previous subsection,  
we take the limit $N,M\to\infty$ with $M=km$, $N=kn$, $k\to\infty$ and $m,n$ kept fixed. 
(As we will see, in terms of QCD with fundamental matter, this means $N_c,N_f\to\infty$ with $N_f/N_c$ fixed.) 
Let us consider the action 
\begin{eqnarray}
S
&=&
S_{gauge}+S_{matter}
=
\int dt\int_{S^3} d^3x\sqrt{g(x)}\left(
{\cal L}_{gauge}+{\cal L}_{matter}
\right), 
\\
{\cal L}_{gauge}
&=&
- \frac{1}{4g_A^2}
TrF_{\mu\nu}^2
- 
\frac{1}{4g_B^2}
TrG_{\mu\nu}^2, 
\\
{\cal L}_{matter}
&=&
k\left(
- Tr (D_\mu\phi_I)(D^\mu\phi_I)^\dagger
+
m_{IJ}Tr \phi_I\phi^\dagger_J
\right),  
\end{eqnarray}
where $F_{\mu\nu}$ and $G_{\mu\nu}$ are field strength of 
$SU(N)$ and $SU(M)$ gauge fields $A_\mu$ and $B_\mu$, 
\begin{eqnarray}
F_{\mu\nu}
=
\partial_\mu A_\nu
-
\partial_\nu A_\mu
-
i[A_\mu,A_\nu], 
\qquad
G_{\mu\nu}
=
\partial_\mu B_\nu
-
\partial_\nu B_\mu
-
i[B_\mu,B_\nu], 
\end{eqnarray}
and $\phi_I (I=1,\cdots,K)$ are $N\times M$ complex matrices 
on which the covariant derivative acts as 
\begin{eqnarray}
D_\mu\phi_I
=
\partial_\mu\phi_I
-
iA_\mu\phi_I
+
i\phi_I B_\mu. 
\end{eqnarray}
We take the mass matrix $m_{IJ}$ to be hermitian. 

In the Maurer-Cartan basis, the matter part of the Lagrangian reads
\begin{eqnarray}
{\cal L}_{matter}
&=&
kTr\Biggl\{
(D_t\phi_I)(D_t\phi_I)^\dagger
- 
(i\mu{\cal L}_i\phi_I-iA_i\phi_I+i\phi_IB_i)
(i\mu{\cal L}_i\phi_I-iA_i\phi_I+i\phi_IB_i)^\dagger 
+ 
m_{IJ}Tr \phi_I\phi^\dagger_J
\Biggl\}. 
\nonumber\\
\end{eqnarray}
The gauge part is the same as (\ref{YM action in Maurer-Cartan basis}) for each gauge group. 

By reducing the spatial dimensions to a point, we obtain 
\begin{eqnarray}
S_{1d}&=&S_{1d}^X+S_{1d}^Y+S_{1d}^{matter}, 
\\
S_{1d}^X
&=&
\frac{C}{g_A^2}\int dt Tr\left(
\frac{1}{2}(D_t X_i)^2
+
\frac{1}{4}[X_i,X_j]^2 
+
i\mu\epsilon_{ijk}X^iX^jX^k 
-
\frac{\mu^2}{2}X_i^2
\right),   
\\
S_{1d}^Y
&=&
\frac{C}{g_B^2}\int dt Tr\left(
\frac{1}{2}(D_t Y_i)^2
+
\frac{1}{4}[Y_i,Y_j]^2 
+
i\mu\epsilon_{ijk}Y^iY^jY^k 
-
\frac{\mu^2}{2}Y_i^2
\right),  
\\
S_{1d}^{matter}
&=&
Ck \int dt
Tr\Biggl\{
(D_t\phi_I)(D_t\phi_I)^\dagger
- 
(X_i\phi_I-\phi_IY_i)
(X_i\phi_I-\phi_IY_i)^\dagger 
+ 
m_{IJ}Tr \phi_I\phi^\dagger_J
\Biggl\} ,   
\end{eqnarray}   
where $X_i$ and $Y_i$ are $N\times N$ and $M\times M$ scalar matrices 
which are obtained from $A_i$ and $B_i$. 
We take the background to be multiple of the background (\ref{IIST background}) 
as in (\ref{multiple IIST background}),    
\begin{eqnarray}
X_i^{b.g.}
=
- \mu L_i\otimes\textbf{1}_n, 
\qquad
Y_i^{b.g.}
=
- \mu L_i\otimes\textbf{1}_m,  
\label{background for quiver}
\end{eqnarray} 
where the size of $L_i$ is $k\times k$. 
Then, it is straightforward to see that the Eguchi-Kawai equivalence works 
around this vacuum. 
By taking $g_{B}^2k$ to be zero, $SU(M)$ reduces to (a part of) 
a global ``flavor'' symmetry.  
The number of flavors turns out to be $N_f = M K$, 
and the ratio $N_f/N_c$ can be arbitrary finite value. 

\subsection{Supersymmetric quiver gauge theory in four dimensions} 
\hspace{0.51cm}
We consider a quiver theory without superpotential. 
(incorporation of the superpotential is straightforward.) 
In this section we use the notation of Wess-Bagger's textbook \cite{WessBagger}. 

The action is given by 
\begin{eqnarray}
S
&=&
S_{gauge}+S_{matter}
=
\int dt\int_{S^3} d^3x\sqrt{g(x)}\left(
{\cal L}_{gauge}+{\cal L}_{matter}
\right), 
\\
{\cal L}_{gauge}
&=&
Tr\left\{
-
\frac{1}{g_A^2}\left(
\frac{1}{4}
F_{\mu\nu}^2
+
i\bar{\lambda}_A\bar{\sigma}^\mu D_\mu\lambda_A
\right)
-
\frac{1}{g_B^2}\left(
\frac{1}{4}
G_{\mu\nu}^2
+
i\bar{\lambda}_B\bar{\sigma}^\mu D_\mu\lambda_B
\right)
\right\}, 
\\
{\cal L}_{matter}
&=&
k Tr\Biggl\{
- 
(D_\mu\phi_I)(D^\mu\phi_I)^\dagger
-
i\bar{\psi_I}\bar{\sigma}^\mu D_\mu\psi_I
-
i\sqrt{2}
\phi^\dagger_I(\lambda_A\psi_I-\psi_I\lambda_B)
\nonumber\\
& &
\qquad
+
i\sqrt{2}
(\bar{\psi}_I\bar{\lambda}_A-\bar{\lambda}_B\bar{\psi}_I)\phi_I
-
\frac{k}{2}
\left(g_A^2\phi_I\phi_I^\dagger\phi_J\phi_J^\dagger
+g_B^2\phi^\dagger_I\phi_I\phi^\dagger_J\phi_J\right)
-
\frac{\mu^2}{8}(\phi_I\phi^\dagger_I)
\Biggl\}, 
\nonumber\\
\end{eqnarray}
where $\psi_I$ and $\phi_I$ belong to $(N,\bar{M})$ representation 
as in the previous subsection. 
(Strictly speaking other multiplets are needed 
in order to cancel the gauge anomaly, 
but we omit them for notational simplicity. 
The modification is straightforward.) 
We notice that the last term is analogous 
to a mass term of adjoint scalars in 4d ${\cal N}=4$ on $S^3$. 

The supersymmetry transformation is 
\begin{eqnarray}
\delta^{(4d)}A_\mu
&=&
-i\bar{\lambda}_A\bar{\sigma}_\mu\epsilon
+
i\bar{\epsilon}\bar{\sigma}_\mu\lambda_A, 
\nonumber\\
\delta^{(4d)}\lambda_A
&=&
F_{\mu\nu}\sigma^{\mu\nu}\epsilon
+
ik g_A^2\phi_I\phi_I^\dagger\epsilon, 
\nonumber\\
\delta^{(4d)}B_\mu
&=&
-i\bar{\lambda}_B\bar{\sigma}_\mu\epsilon
+
i\bar{\epsilon}\bar{\sigma}_\mu\lambda_B, 
\nonumber\\
\delta^{(4d)}\lambda_B
&=&
G_{\mu\nu}\sigma^{\mu\nu}\epsilon
+
ik g_B^2\phi^\dagger_I\phi_I\epsilon,
\nonumber\\
\delta^{(4d)}\phi_I
&=&
\sqrt{2}\epsilon\psi_I, 
\nonumber\\ 
\delta^{(4d)}\psi_I
&=&
i\sqrt{2}\sigma^\mu\bar{\epsilon}D_\mu\phi_I
+
\frac{\mu}{4\sqrt{2}}\bar{\epsilon}\phi_I. 
\end{eqnarray}
Here the supersymmetry transformation parameter $\epsilon$ satisfies 
\begin{eqnarray}
D_\mu\epsilon
=
-\frac{i\mu}{4}
\sigma_\mu\epsilon.   
\end{eqnarray} 
 
The dimensionally reduced model is obtained by rewriting the action 
in the Maurer-Cartan basis and then by reducing the spatial dimensions.  
It is important that the parameters of the supersymmetry transformation 
depends only on $t$ in this basis
\begin{eqnarray}
\epsilon(t)
=
e^{-i\mu t/4}\epsilon_0 . 
\end{eqnarray}
Consequently the dimensional reduction of spatial dimensions 
does not affect supersymmetry.  

By dimensionally reducing the spatial directions we obtain the matrix quantum mechanics
\begin{eqnarray}
L_{gauge}^{m.m.}
&=&
\frac{C}{g_A^2}
Tr
\Biggl(
\frac{1}{2}
\left(
D_t X_i 
\right)^2 
+
\frac{1}{4}[X_i,X_j]^2
-
\frac{\mu^2}{2}X_i^2
+
i\mu\epsilon^{ijk}X_iX_jX_k 
\Biggl) 
\nonumber\\
& &
+
\frac{C}{g_A^{2}}
Tr\left(
-
i\bar{\lambda}_XD_t\lambda_X
-
\bar{\lambda}_X\bar{\sigma}^i
[X_i,\lambda_X]
-
\frac{3}{4}
\mu\bar{\lambda}_X\lambda_X
\right)
\nonumber\\
& &
+
(X\to Y), 
\end{eqnarray}
\begin{eqnarray}
L_{matter}^{m.m.}
&=&
Ck Tr\Biggl\{
(D_t\phi_I)(D_t\phi_I)^\dagger
- 
(X_i\phi_I-\phi_IY_i)
(X_i\phi_I-\phi_IY_i)^\dagger
\nonumber\\
& &
\qquad
-
i\bar{\psi}_ID_t\psi_I
-
\bar{\psi}_I\bar{\sigma}^i
(X_i\psi_I-\psi_I Y_i)
-
\frac{3}{4}
\mu\bar{\psi}_I\psi_I 
\nonumber\\
& &
\qquad
-
i\sqrt{2}
\phi^\dagger_I(\lambda_X\psi_I-\psi_I\lambda_Y)
+
i\sqrt{2}
(\bar{\psi}_I\bar{\lambda}_X-\bar{\lambda}_Y\bar{\psi}_I)\phi_I
\nonumber\\
& &
\qquad
-
\frac{k}{2}
\left(g_A^2\phi_I\phi_I^\dagger\phi_J\phi_J^\dagger
+g_B^2\phi^\dagger_I\phi_I\phi^\dagger_J\phi_J\right)
-
\frac{\mu^2}{8}(\phi_I\phi^\dagger_I)
\Biggl\}. 
\nonumber\\
\end{eqnarray} 

The supersymmetry transformation reduces to 
\begin{eqnarray}
\delta^{(1d)}X_i
&=&
-i\bar{\lambda}_X\bar{\sigma}_i\epsilon
+
i\bar{\epsilon}\bar{\sigma}_i\lambda_X, 
\nonumber\\
\delta^{(1d)}A_t
&=&
-i\bar{\lambda}_X\epsilon+i\bar{\epsilon}\lambda_X,  
\nonumber\\
\delta^{(1d)}\lambda_X
&=&
\left[
-2(D_tX_i)
+
\epsilon^{ijk}[X_j,X_k]
+
2i\mu X_i
\right]\sigma^i\epsilon
+
ik g_A^2\phi_I\phi_I^\dagger\epsilon, 
\nonumber\\
\delta^{(1d)}Y_i
&=&
-i\bar{\lambda}_Y\bar{\sigma}_i\epsilon
+
i\bar{\epsilon}\bar{\sigma}_i\lambda_Y, 
\nonumber\\
\delta^{(1d)}B_t
&=&
-i\bar{\lambda}_Y\epsilon+i\bar{\epsilon}\lambda_Y, 
\nonumber\\
\delta^{(1d)}\lambda_B
&=&
\left[
-2(D_tY_i)
+
\epsilon^{ijk}[Y_j,Y_k]
+
2i\mu Y_i
\right]\sigma^i\epsilon
+
ik g_B^2\phi^\dagger_I\phi_I\epsilon,
\nonumber\\
\delta^{(1d)}\phi_I
&=&
\sqrt{2}\epsilon\psi_I, 
\nonumber\\ 
\delta^{(1d)}\psi_I
&=&
\sqrt{2}\sigma^i\bar{\epsilon}(X_i\phi_I-\phi_IY_i)
+
\frac{\mu}{4\sqrt{2}}\bar{\epsilon}\phi_I,  
\end{eqnarray}
where $\epsilon$ is time-dependent
\begin{eqnarray}
\epsilon(t)
=
e^{-i\mu t/4}\epsilon_0. 
\end{eqnarray}

It turns out that the background (\ref{background for quiver}) preserve this supersymmetry. 
By expanding the matrix model around this background we recover the 
original 4d theory.   
Notice that we have to renormalize the bare 't Hooft couplings 
$g_A^2N$ and $g_B^2M$ appropriately in the continuum limit.  

Generalizations to more complicated quiver theories are straightforward. 
We emphasize that the equivalence works only when the field theory  
does not has gauge anomalies, because we assumed implicitly in the proof.  

The emergence of the chiral anomaly is a long standing problem 
in Eguchi-Kawai models. This problem exists 
in the present case too -- the chiral symmetry seems kept in 
the reduced model. 
The chiral symmetry should be broken by some effect. 
Here, we do not pursue this direction, but 
just assume the presence of such an effect. 
One possible way to find it 
is to consider the chiral anomaly in the noncommutative 
space (concentric fuzzy spheres). 

\section{The Eguchi-Kawai model for Chern-Simons gauge theories}\label{section:CS}
\hspace{0.51cm}
In this section, we consider the Eguchi-Kawai model 
of three dimensional Chern-Simons gauge theories. 
In a similar fashion to the YM cases, 
we can obtain matrix models which have 
a fuzzy sphere as a classical solution. 
Quivers can be introduced into this construction 
as in the previous section. 
We consider the ABJM model \cite{ABJM08} as an example. 
\subsection{Bosonic case} 
\hspace{0.51cm}
Let us start with the bosonic Chern-Simons with $U(N)$ gauge group. 
The action is 
\begin{eqnarray}
S_{CS}
&=&
i\cdot \frac{k}{4\pi}\int d^3x\sqrt{g(x)} \epsilon^{\mu\nu\rho} Tr\left(
-A_\mu \partial_\nu A_\rho
-
\frac{2i}{3}A_\mu A_\nu A_\rho
\right), 
\end{eqnarray}
where $k$ is an integer. On the three-sphere of radius $2/\mu$, 
by using the Maurer-Cartan basis, this action can be written as \cite{IIOST07}
\begin{eqnarray}
S_{CS}
&=&  
i\cdot \frac{k}{4\pi}\left(\frac{2}{\mu}\right)^3\int d^3\Omega  Tr\left[
\epsilon^{ijk}
\left(
\frac{i\mu}{2}
\left(
-
({\cal L}_i A_j)A_k
+
({\cal L}_j A_i)A_k
\right)
-
\frac{2i}{3}A_i A_j A_k
\right)
+
\mu A_i^2
\right]. 
\nonumber\\
\label{CS_3d action}
\end{eqnarray}
Dimensionally reducing it, we obtain 
\begin{eqnarray}
S_{mm}
=
iC\cdot \frac{k}{4\pi}
Tr\left(
-\frac{2i}{3}\epsilon^{\mu\nu\rho} 
X_\mu X_\nu X_\rho
+
\mu X_i^2\right). 
\label{CS_0d action}
\end{eqnarray}
This theory has a classical solution 
\begin{eqnarray}
X_i= - \mu L_i, 
\qquad
[L_i,L_j]=i\epsilon_{ijk}L_k. 
\end{eqnarray}

It is easy to check that the 3d action (\ref{CS_3d action}) is obtained 
from (\ref{CS_0d action}) by taking the background (\ref{IIST background}) 
and using the mapping rule given in \S~\ref{sec:3d example}. 
\subsection{ABJM theory} 
\hspace{0.51cm}
The previous construction can be easily promoted 
to supersymmetric Chern-Simons theory. 
As a concrete example, let us formulate the ABJM model, 
which gives a description of the multiple M2-branes theory \cite{ABJM08} (see also \cite{HLLP08}). 
In the following we take the gauge group to be $U(N)\times U(M)$. 
When this model is put on the three sphere, 
an additional mass term must be added in order to 
keep supersymmetry. 
The model is given by 
\begin{eqnarray}
{\cal L}
&=&
\frac{ik}{4\pi}\epsilon^{\mu\nu\rho}
Tr
\left(
-A_\mu \partial_\nu A_\rho
-
\frac{2i}{3}A_\mu A_\nu A_\rho
+
B_\mu \partial_\nu B_\rho
+
\frac{2i}{3}B_\mu B_\nu B_\rho
\right)
\nonumber\\
& &
+
\frac{k}{2\pi}
Tr\left(
D_\mu\bar{\phi}^\alpha D^\mu\phi_\alpha
+
i\bar{\psi}_\alpha\sigma^\mu D_\mu\psi^\alpha
\right)
+
k{\cal V}(\phi,\psi), 
\end{eqnarray}
where
\begin{eqnarray}
{\cal V}(\phi,\psi)
&=&
-\frac{i}{2\pi}
\epsilon^{\alpha\beta\gamma\delta}
Tr\left(
\phi_\alpha\bar{\psi}_\beta\phi_\gamma\bar{\psi}_\delta
\right)
+
\frac{i}{2\pi}
\epsilon_{\alpha\beta\gamma\delta}
Tr\left(
\bar{\phi}^\alpha\psi^\beta\bar{\phi}^\gamma\psi^\delta
\right)
\nonumber\\
& &
-
\frac{i}{2\pi}
 Tr\left(
\bar{\phi}^\alpha\phi_\alpha\bar{\psi}_\beta\psi^\beta
-
\phi_\alpha\bar{\phi}^\alpha\psi^\beta\bar{\psi}_\beta
+
2\bar{\phi}^\alpha\psi^\beta\bar{\psi}_\alpha\phi_\beta
-
2\phi_\alpha\bar{\psi}_\beta\psi^\alpha\bar{\phi}^\beta
\right)
\nonumber\\
& &
-
\frac{1}{6\pi}
Tr\left(
(\phi_\alpha\bar{\phi}^\alpha)^3
+
(\bar{\phi}^\alpha\phi_\alpha)^3
\right)
-
\frac{2}{3\pi}
Tr\left(
\phi_\alpha\bar{\phi}^\gamma
\phi_\beta\bar{\phi}^\alpha
\phi_\gamma\bar{\phi}^\beta
\right)
\nonumber\\
& &
+
\frac{1}{\pi}
Tr\left(
\phi_\alpha\bar{\phi}^\alpha
\phi_\beta\bar{\phi}^\gamma
\phi_\gamma\bar{\phi}^\beta
\right) 
+
\frac{3\mu^2}{32\pi}
\bar{\phi}^\alpha\phi_\alpha. 
\end{eqnarray} 
Here $A_\mu$ and $B_\mu$ are gauge fields of $U(N)$ and $U(M)$ groups, respectively. 
Bifundamental matter fields $\phi_\alpha,\psi^\alpha$ ($\alpha=1,\cdots,4$) 
are in the $(N,\bar{M})$ representations and the covariant derivative $D_\mu$ acts as 
\begin{eqnarray}
D_\mu\phi_\alpha
=
\nabla_\mu\phi_\alpha
-
iA_\mu\phi_\alpha
+
i\phi_\alpha B_\mu, 
\end{eqnarray}
and similarly for $\psi^\alpha$. 
We have rescaled these matter fields in such a way that 
the Chern-Simons level $k$ appears only 
in the overall factor. 
 
The supersymmetry transformation is 
\begin{eqnarray}
\delta^{(3d)}\phi_\alpha
&=&
-i\eta_{\alpha\beta}\psi^\beta, 
\nonumber\\ 
\delta^{(3d)}A_\mu
&=&
-\left(
\eta^{\alpha\beta}\sigma_\mu\phi_\alpha\bar{\psi}_\beta
+
\eta_{\alpha\beta}\sigma_\mu\psi^\beta\bar{\phi}^\alpha
\right), 
\nonumber\\
\delta^{(3d)}B_\mu
&=&
-\left(
\eta^{\alpha\beta}\sigma_\mu\bar{\psi}_\beta\phi_\alpha
+
\eta_{\alpha\beta}\sigma_\mu\bar{\phi}^\alpha\psi^\beta
\right), 
\nonumber\\
\delta^{(3d)}\psi^\alpha
&=&
\left[
\sigma^\mu D_\mu\phi_\gamma
-
\frac{2}{3}\phi_{[\beta}\bar{\phi}^\beta\phi_{\gamma]}
\right]
\eta^{\gamma\alpha}
+
\frac{4}{3}
\phi_{\beta}\bar{\phi}^\alpha\phi_{\gamma}\eta^{\gamma\beta}
+
\frac{2}{3}\epsilon^{\alpha\beta\gamma\delta}
\phi_{\beta}\bar{\phi}^\rho\phi_{\gamma}
\eta_{\delta\rho}
-
\frac{i\mu}{4}\eta^{\gamma\alpha}\phi_\gamma ,     
\nonumber\\
\end{eqnarray} 
where the parameter satisfies $\eta^{\alpha\beta}=-\eta^{\beta\alpha}$, 
$(\eta_{\alpha\beta})^\ast
=
\frac{1}{2}\eta^{\alpha\beta\gamma\delta}\eta_{\gamma\delta}
=\eta^{\alpha\beta}$ 
and  
$\nabla_\mu\eta^{\alpha\beta}
=
-\frac{i\mu}{4}\sigma^\mu
\eta^{\alpha\beta}$.   

Rewriting the action using the Maurer-Cartan basis and taking the zero-dimensional reduction, 
we obtain the matrix model 
\begin{eqnarray}
S_{mm}
&=&
Ck\biggl[
\frac{i}{4\pi}
Tr
\left(
-\frac{2i}{3}\epsilon^{ijk} 
X_i X_j X_k
+
\mu X_i^2
+
\frac{2i}{3}\epsilon^{ijk} 
Y_i Y_j Y_k
-
\mu Y_i^2
\right)
\nonumber\\
& &
-
\frac{1}{2\pi}
Tr
(Y_i\bar{\phi}^\alpha-\bar{\phi}^\alpha X_i) 
(X^i\phi_\alpha-\phi_\alpha Y^i)
\nonumber\\
& &
+
\frac{1}{2\pi}
Tr\left(\bar{\psi}_\alpha\sigma^i 
\left(
X_i\psi^\alpha
-
\psi^\alpha Y_i
\right)
+
\frac{3\mu}{4}\bar{\psi}_\alpha\psi^\alpha
\right)
+
{\cal V}(\phi,\psi)\biggl]. 
\end{eqnarray} 
Expanding this action around the background (\ref{IIST background}), 
we can reproduce the original action. 
As before, the transformation parameter in the 3d theory 
is a constant in the Maurer-Cartan basis 
and the reduced model is supersymmetric. 
The supersymmetry transformation is 
\begin{eqnarray}
\delta^{(0d)}\phi_\alpha
&=&
-i\eta_{\alpha\beta}\psi^\beta, 
\nonumber\\ 
\delta^{(0d)}X_i
&=&
-\left(
\eta^{\alpha\beta}\sigma_i\phi_\alpha\bar{\psi}_\beta
+
\eta_{\alpha\beta}\sigma_i\psi^\beta\bar{\phi}^\alpha
\right), 
\nonumber\\
\delta^{(0d)}Y_i
&=&
-\left(
\eta^{\alpha\beta}\sigma_i\bar{\psi}_\beta\phi_\alpha
+
\eta_{\alpha\beta}\sigma_i\bar{\phi}^\alpha\psi^\beta
\right), 
\nonumber\\
\delta^{(0d)}\psi^\alpha
&=&
\left[
\sigma^i (-iX_i\phi_\gamma+i\phi_\gamma Y_i)
-
\frac{2}{3}\phi_{[\beta}\bar{\phi}^\beta\phi_{\gamma]}
\right]
\eta^{\gamma\alpha}
+
\frac{4}{3}
\phi_{\beta}\bar{\phi}^\alpha\phi_{\gamma}\eta^{\gamma\beta}
+
\frac{2}{3}\epsilon^{\alpha\beta\gamma\delta}
\phi_{\beta}\bar{\phi}^\rho\phi_{\gamma}
\eta_{\delta\rho}
-
\frac{i\mu}{4}\eta^{\gamma\alpha}\phi_\gamma.
\nonumber\\
\end{eqnarray}
Notice that this background preserves 
all supersymmetries. 

In the actions the Chern-Simons level $k$ appears as an overall factor. 
The original ABJM is reproduced from the reduced model in the planar limit
with  both $k/N$ and $k/M$ kept fixed. 
 
\section{Conclusion and Discussions}
\hspace{0.51cm}
In this paper we have applied a recently proposed large-$N$ reduction 
technique \cite{IIST08} to supersymmetric quiver and Chern-Simons theories.  
As concrete examples we have constructed 
the $SU(N)\times SU(M)$ supersymmetric quiver gauge theory 
with bifundamental matter fields 
and the ABJM model of multiple M2 branes. 
Furthermore, by taking one of the gauge couplings to be small 
in the supersymmetric quiver gauge theory 
we obtain $SU(N)$ supersymmetric QCD with flavor. 
In this construction both $N_c$ and $N_f$ are infinite 
but the ratio $N_f/N_c$ can take any value. 
Therefore this construction provides us with a valuable tool to study 
the dynamics of supersymmetric QCD, e.g. supersymmetry breaking, 
Seiberg duality conjecture \cite{Seiberg94}, etc. 

The reduced model of the ABJM theory will be useful 
to study the AdS/CFT correspondence numerically. 
Of particular interest is the strong 't Hooft coupling region 
that is expected to describe 
type IIA string on $AdS^4\times {\mathbb CP}^3$. 
This region can be studied by using the Eguchi-Kawai equivalence. 
It turns out that the parameter region where $k$ is smaller than $O(N)$ 
is also important to obtain insights into M-theory. 
Thermodynamical properties are also interesting. 
For these reasons it is still valuable to study a lattice formulation 
which is valid at finite-$N$ and can be put at finite temperature. 
For references in this direction, see e.g. \cite{ChernSimons}.

We expect that these models have the sign problem, 
unlikely to the reduced model for 4d ${\cal N}=1$ pure SYM \cite{HMM09}. 
This problem possibly make it difficult to study these models numerically. 
At finite temperature, however, the sign problem might be mild, 
similarly to the case of the maximally supersymmetric 
matrix quantum mechanics \cite{AHNT07}. 
It is important to study by direct simulation 
how severe the sign problem is. 

Although inherently restricted to the planar limit, the Eguchi-Kawai equivalence 
can be a powerful tool to explore the dynamics of supersymmetric gauge theories. 
Numerical studies on these models will be reported 
in future communications.

\section*{Acknowledgments}
\hspace{0.51cm}
The authors are grateful to G.~Ishiki, J.~Nishimura and H.~Suzuki 
for stimulating discussions and comments. 
The authors thank the Yukawa Institute for 
Theoretical Physics at Kyoto University and 
participants of the YITP workshop YITP-W-09-04 on 
``Development of Quantum Field Theory and String Theory'' 
who gave us useful comments. 


\end{document}